\documentclass[aps,twocolumn,nofootinbib,preprintnumbers,groupedaddress,eqsecnum]{revtex4-1}
\usepackage[T1]{fontenc}
\usepackage[latin9]{inputenc}
\usepackage{geometry}
\usepackage{verbatim}
\usepackage{amssymb}
\usepackage{setspace}

\usepackage{amsmath}
\usepackage{graphicx, xcolor, varwidth}
\usepackage{hyperref}
\usepackage{pdfpages}
\usepackage{subcaption}

\makeatletter
\AtBeginDocument{\let\LS@rot\@undefined}
\makeatother

\begin{document}

\title{Mutation rate variability as a driving force in adaptive evolution}

\author{Dalit Engelhardt}
\email{dengelhardt@fas.harvard.edu}
\author{Eugene I. Shakhnovich}
\email{shakhnovich@chemistry.harvard.edu}

\affiliation{Department of Chemistry and Chemical Biology, Harvard University, Cambridge, MA 02138}

\begin{abstract}
Mutation rate is a key determinant of the pace as well as outcome of evolution, and variability in this rate has been shown in different scenarios to play a key role in evolutionary adaptation and resistance evolution under stress caused by selective pressure. Here we investigate the dynamics of resistance fixation in a bacterial population with variable mutation rates and show that evolutionary outcomes are most sensitive to mutation rate variations when the population is subject to environmental and demographic conditions that suppress the evolutionary advantage of high-fitness subpopulations. By directly mapping a biophysical fitness function to the system-level dynamics of the population we show that both low and very high, but not intermediate, levels of stress in the form of an antibiotic result in a disproportionate effect of hypermutation on resistance fixation. We demonstrate how this behavior is directly tied to the extent of genetic hitchhiking in the system, the propagation of high-mutation rate cells through association with high-fitness mutations. Our results indicate a substantial role for mutation rate flexibility in the evolution of antibiotic resistance under conditions that present a weak advantage over wildtype to resistant cells.
\end{abstract}

\maketitle

\section{Introduction}
The ability to predict the possible trajectories of a naturally evolving complex living system is key to describing and anticipating varied ecological and biomedical phenomena. Such predictability rests on an understanding of the \textit{potential for evolutionary adaptability} of a given system.  In asexual populations a major mechanism responsible for evolutionary adaptation under environmental stress is the generation via genetic mutations of phenotypes able to better withstand and thrive under the stressor: resistant populations arising from within a wildtype population that may ``rescue'' the population from the source of stress by eventually coming to dominate the population. The rate at which such resistant mutations occur and the balance between these and more deleterious mutations are major determinants of whether the population may survive and adapt to selective evolutionary pressure~\cite{lynch2016genetic,denamur2006evolution,de2002fate,sniegowski2000evolution,taddei1997role}, an environmental stressor that targets strain variants, or phenotypes, non-uniformly. Although the baseline mutation rate in bacteria is quite low, at about $\sim 10^{-3}$ per genome per generation~\cite{lee2012rate,drake1998rates}, high prevalences of mutator strains in natural bacterial populations and clinical isolates have been observed in various studies (see~\cite{gross1981incidence,leclerc1996high,matic1997highly,denamur2002high} for early work and~\cite{hall2006hypermutable} for a survey), and in certain cases ``hypermutability'', an increase in the mutation rate over the baseline rate, was shown to result in fitness increases and faster adaptation~\cite{chao1983competition,taddei1997role,tenaillon1999mutators,heo2010interplay,ram2012evolution,ram2014stress,lukavcivsinova2017stress} and even be essential for survival under stress~\cite{swings2017adaptive} by enabling genetic hitchhiking on beneficial mutations~\cite{smith1974hitch,taddei1997role,gentile2011competition,giraud2001costs}. Mutation rates can increase under environmental stress~\cite{fitzgerald2017stress,maclean2013evaluating,galhardo2007mutation,foster2007stress}, and, in particular, hypermutability may play a significant role in the rise of antibiotic resistance~\cite{hammerstrom2015acinetobacter,jolivet2011bacterial,eliopoulos2003hypermutation,chopra2003role,giraud2002mutator,martinez2000mutation}.

The potential for adaptability via genetic mutations is dependent on the interplay between the ensemble of phenotypes that the system can access via mutations and the rate at which such transitions may occur within this ensemble. Phenotypes are typically characterized by some \textit{intrinsic} measure of evolutionary fitness, such as their growth rate or lag phase, that contributes to evolutionary success, with \textit{extrinsic} conditions, such as the probability of acquisition of this trait, initial population distribution, or resource availability, held fixed. Yet evolutionary advantage is determined by an interplay of these intrinsic and extrinsic factors, and separating these dependences while considering only a subset of them is of limited utility in establishing a global picture of a system's evolvability potential as well as specific response to selective pressure. Here, we address both with a view to investigating the extent to which mutation rate variability drives adaptation under selective pressure.

The main purpose of this paper is to demonstrate that evolution under selective pressure -- an external stressor that induces a fitness gradient in a given population -- may not be uniformly sensitive to mutation rate as a function of the selective pressure as well as additional fitness-determining conditions, and that this non-uniform behavior should be taken into account when deciding on an appropriate antibiotic dosing protocol. In such a situation there is generally no information available on the mutation rate in the pathogenic bacterial population, and this rate may also change in the course of therapy, as noted above. If dosing can be restricted to ranges for which the expected evolutionary outcome is less sensitive to the mutation rate, there will be higher predictive certainty about the treatment outcome and more reliable strategies can be developed for avoiding antibiotic resistance arising in the course of treatment.

By considering a simple deterministic model of bacterial evolution under limited resources, we show that evolutionary outcome is most sensitive to the mutation rate when there exist phenotypes in the population that have a weak advantage -- expressed through either intrinsic traits or extrinsic conditions -- over the phenotype that is initially dominant in the population. In Section~\ref{sec:model} we introduce and describe our evolutionary dynamics model; in Section~\ref{sec:advantage} we define and motivate our measure of mutation rate sensitivity and quantify how sensitive the evolutionary success of a population is to increases in the mutation rate. We show that the fitness advantage of the resistant mutant -- as given by both intrinsic fitness and extrinsic advantage-conferring conditions -- is a determining factor in the extent of this sensitivity. In Section~\ref{sec:inhibitioneffects}, we focus our analysis on selective evolutionary pressure in the form of a bacterial growth inhibitor (antibiotic) and quantify (i) the dependence of mutation rate sensitivity to this source of pressure, and (ii) the extent to which the antibiotic drives the fixation of hypermutation in the population. We conclude with a brief discussion of the ramifications of our findings in Section~\ref{sec:discussion}.

\section{The Model}\label{sec:model}

We consider a system under non-neutral selection in which up to two distinct phenotypes -- defined by their growth rates $g_i$ -- may coexist under limited resources. We denote them by $wt$ for wildtype and $r$ for resistant -- designations that are intended to indicate that the $r$ phenotype is more resistant to the evolutionary pressure considered (e.g. antibiotic, as in Section~\ref{sec:inhibitioneffects}) under the non-neutral selection experienced by the population. Each of these phenotypes may be found with some baseline mutation rate $\mu_{bl}$ or with an elevated mutation rate of $f\times\mu_{bl}$, $f>1$. Both forward and backward mutations are permitted with equal probability\footnote{While resistance can result from the accumulation of a series of mutations, first-generation mutants -- for which the mutation is reversible by a single-point mutation -- can already exhibit discernibly increased resistance~\cite{palmer2015delayed, rodrigues2016biophysical}, with further substantial increases in resistance found in some second-generation mutants, for which the equal probability assumption can be thought of as a ``first-order'' approximation.} $p_{wt,r}$; transitions between baseline-mutation phenotype and its elevated-mutation counterpart (of identical growth rate) occur with rate $r_{\mu}$. In addition, to account for the fitness penalty incurred due to an increased rate of deleterious mutations at higher values of $f$, we assume that either phenotype may experience deleterious mutations with probability $P_{del}$. Since at non-negligible levels of selective pressure phenotypes whose resistance to the pressure is weaker than wildtype will have very low growth, we do not keep track of such low-growth populations explicitly, but they are implicitly accounted for in our model as the loss of cells from higher-growth populations via deleterious mutations. 
\begin{figure}[t]
	\centering
	\includegraphics[height=1.4in]{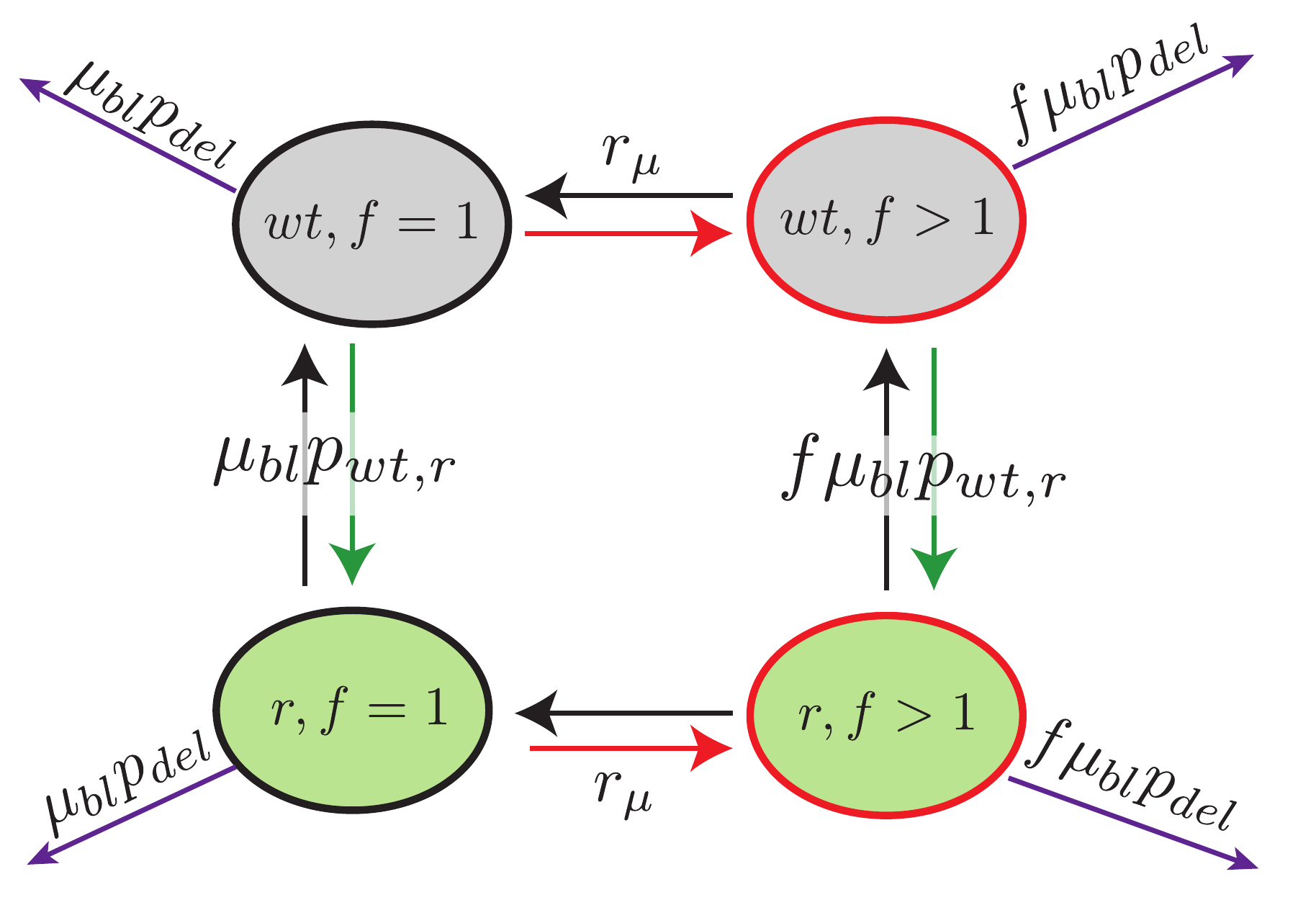}
	\caption{\footnotesize{Schematic indicating the allowed single-step transitions and their rates between phenotypes.}}
	\label{fig:schematic}
\end{figure}
Since we assume that such loss occurs with uniform probability $P_{del}$, when the overall genetic mutation rate of a cell is $\mu$, the rate of deleterious mutations is given by $\mu P_{del}$. Note that when $P_{del}$ is high, increases in $\mu$ carry a higher penalty, implying that for hypermutation to be beneficial and counteract this penalty resistant phenotypes would have to be significantly advantageous either by having a much higher growth rate (intrinsic advantage) or, e.g., by occurring with a high probability or being initially present in relatively high proportions (extrinsic advantage). Fig.~\ref{fig:schematic} shows a schematic od this model.  We assume deterministic evolution under limited resources, as resources needed for growth are nearly always constrained in real biological systems, driving competition between organisms consuming the same resources. The equations governing the time evolution of this system are given by
\begin{equation}
\begin{aligned}\dot{x}_{k,\alpha}(t)= & \left(1-\frac{x_{tot}(t)}{K}\right)\left[g_{k}x_{k,\alpha}(t)\right.\\
& +\mu_{bl}f_{\alpha}p_{j,k}g_{j}x_{j,\alpha}(t)+r_{\mu}g_{k}x_{k,\beta}(t)\\
& \left.-f_{\alpha}\mu_{bl}g_k\left(p_{k,j}+P_{del}\right)-r_{\mu}g_{k}x_{k,\alpha}(t)\right]
\end{aligned}
\label{eq:gensystem}
\end{equation}
where $j\neq k$, $j,k\in\{wt,r\}$, $\alpha\neq\beta$, and a stationary population distribution is established when the total population size $x_{tot}=\sum_{m,\gamma}x_{m,\gamma}$ reaches the resource capacity $K$. Note that faster-growing phenotypes will also produce exponentially more deleterious mutants as a result of their more frequent divisions, resulting in the previously noted fitness tradeoff. The four-dimensional system (Fig.~\ref{fig:schematic}) of Eqn. (\ref{eq:gensystem}) is given explicitly in Appendix~\ref{sec:eveqns}. We consider here the case where the effect of selective pressure is limited to \textit{selecting} for hypermutant variants if they are advantageous but not directly inducing hypermutability (for work on the latter see e.g.,~\cite{phillips1987induction,piddock1987induction,ysern1990induction,ren1999escherichia,perez2005sos}). Under this assumption, hypermutation occurs independently of selective pressure and therefore  some proportion of the initial-state population would be expected to already exhibis elevated mutation rates. We assume in all that follows that cells with elevated mutation rates constitute $1\%$ of the total initial population (distributed in proportion to the phenotype distribution) and a corresponding rate at which hypermutation-conferring mutations occur of $0.25\%$ of cells per generation (see Appendix~\ref{sec:rmut} for an extended discussion of these parameter choices).

\section{Sensitivity of evolutionary success to the mutation rate}\label{sec:advantage}

\begin{figure*}[htb!]
	\begin{center}
		\begin{subfigure}[t]{0.24\linewidth}
			\includegraphics[width=1\linewidth]{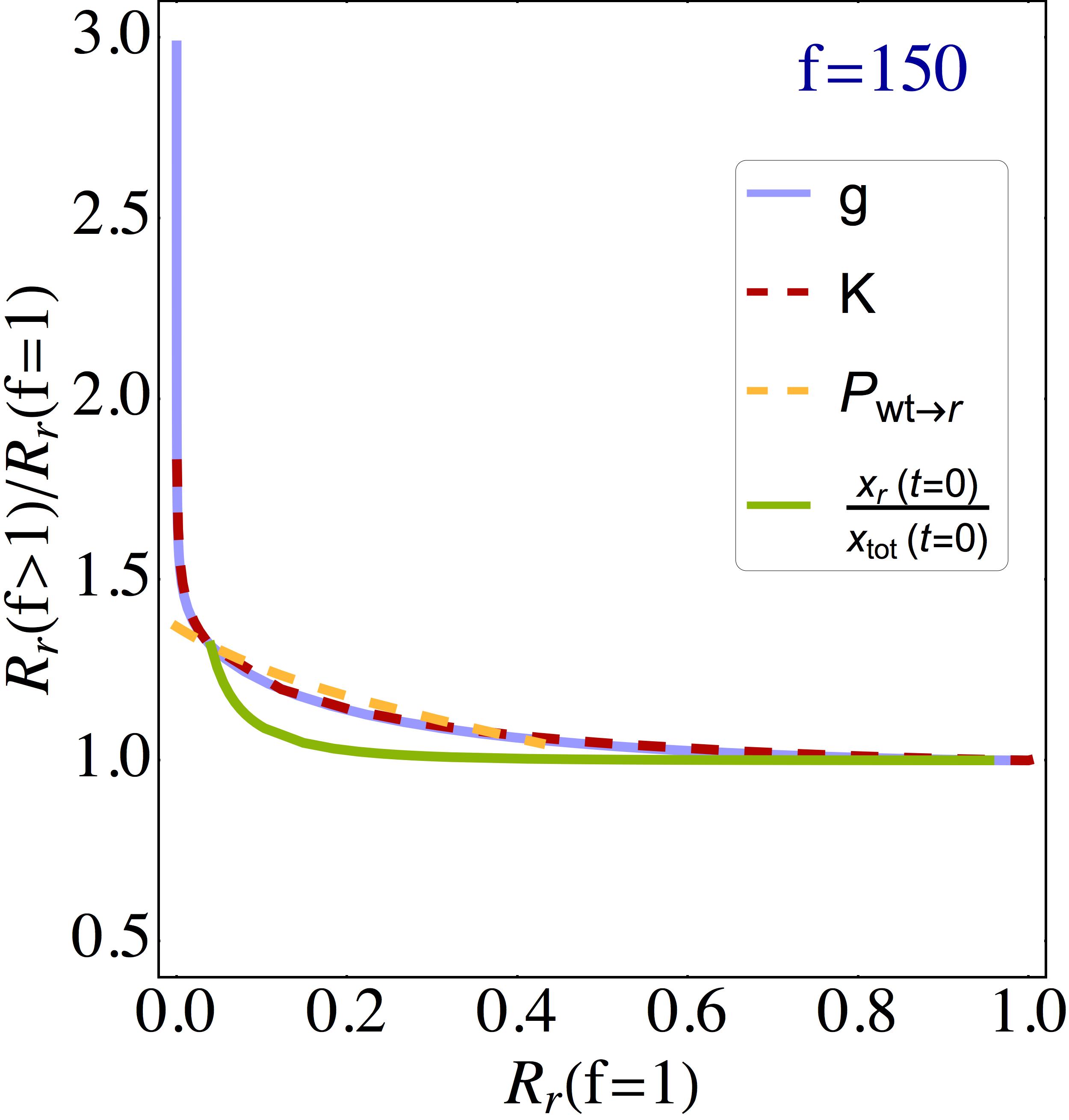} 
			\caption{}
			\label{subfig:singleRr}
		\end{subfigure}
		\begin{subfigure}[t]{0.24\linewidth}
			\includegraphics[width=1\linewidth]{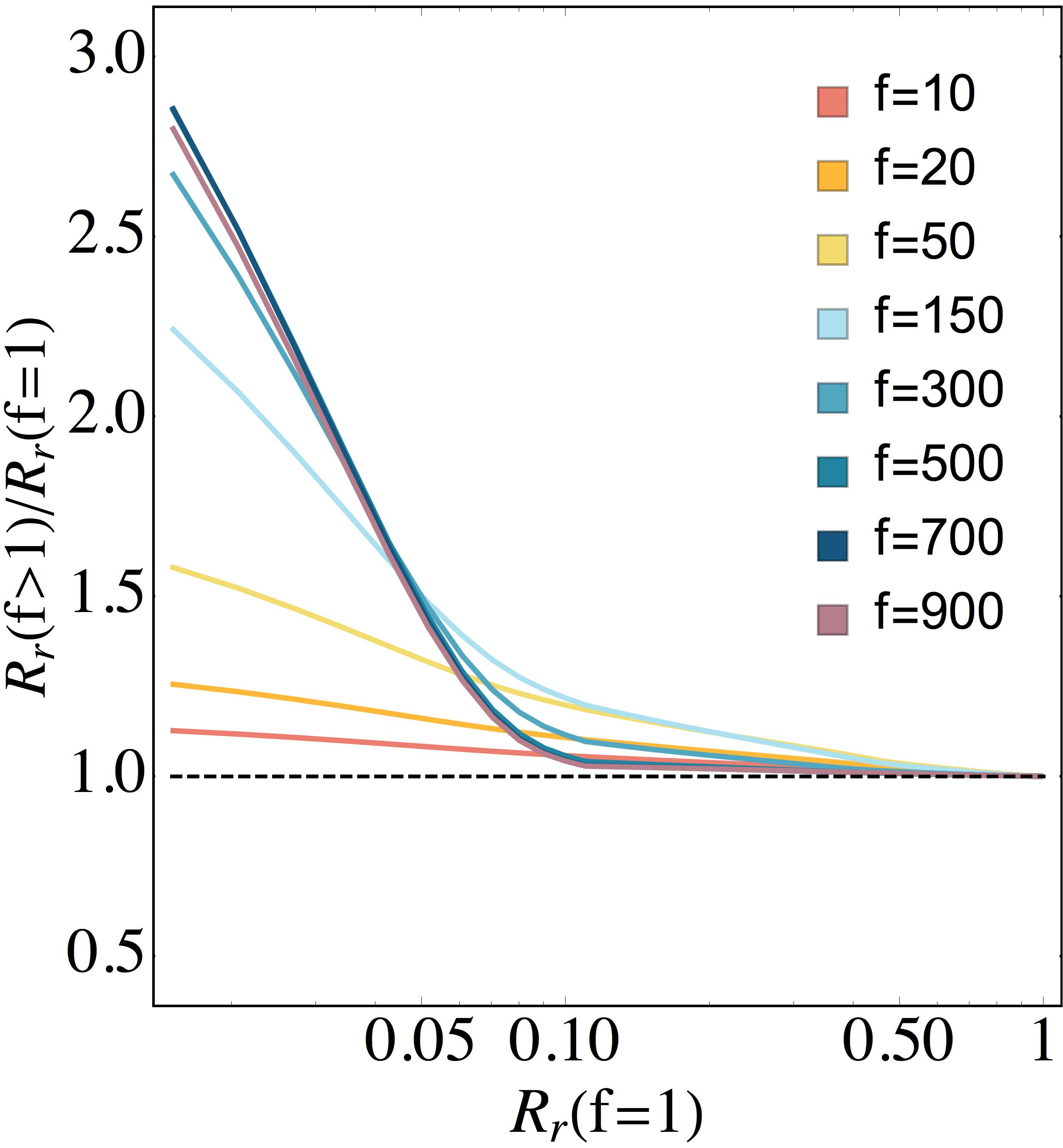} 
			\caption{}
			\label{subfig:aveRr}
		\end{subfigure}\hfill
		\begin{subfigure}[t]{0.24\linewidth}
			\includegraphics[width=1\linewidth]{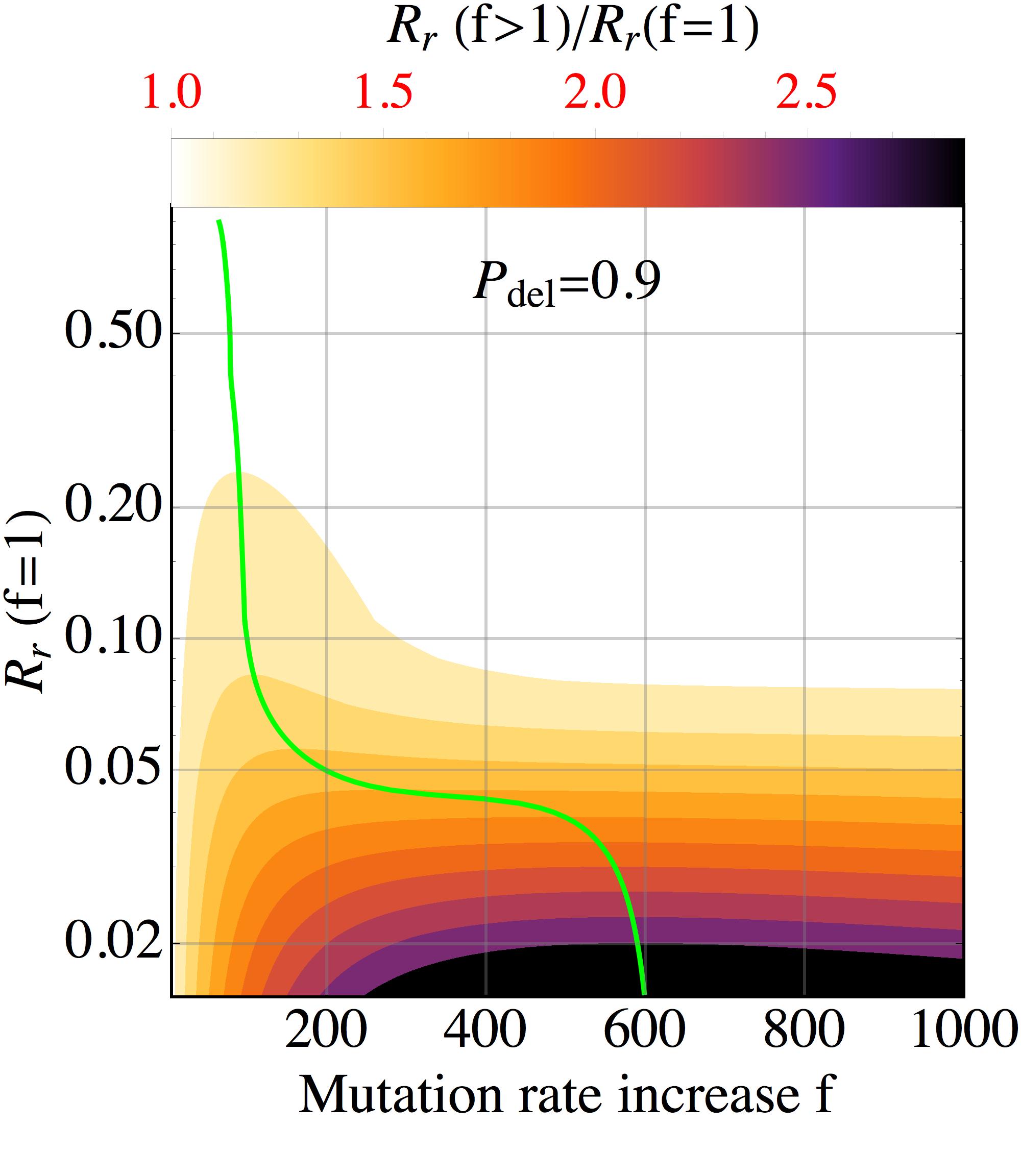} 
			\caption{}
			\label{subfig:pdel09}
		\end{subfigure}\hspace{-0.1cm}
		\begin{subfigure}[t]{0.24\linewidth}
			\includegraphics[width=1\linewidth]{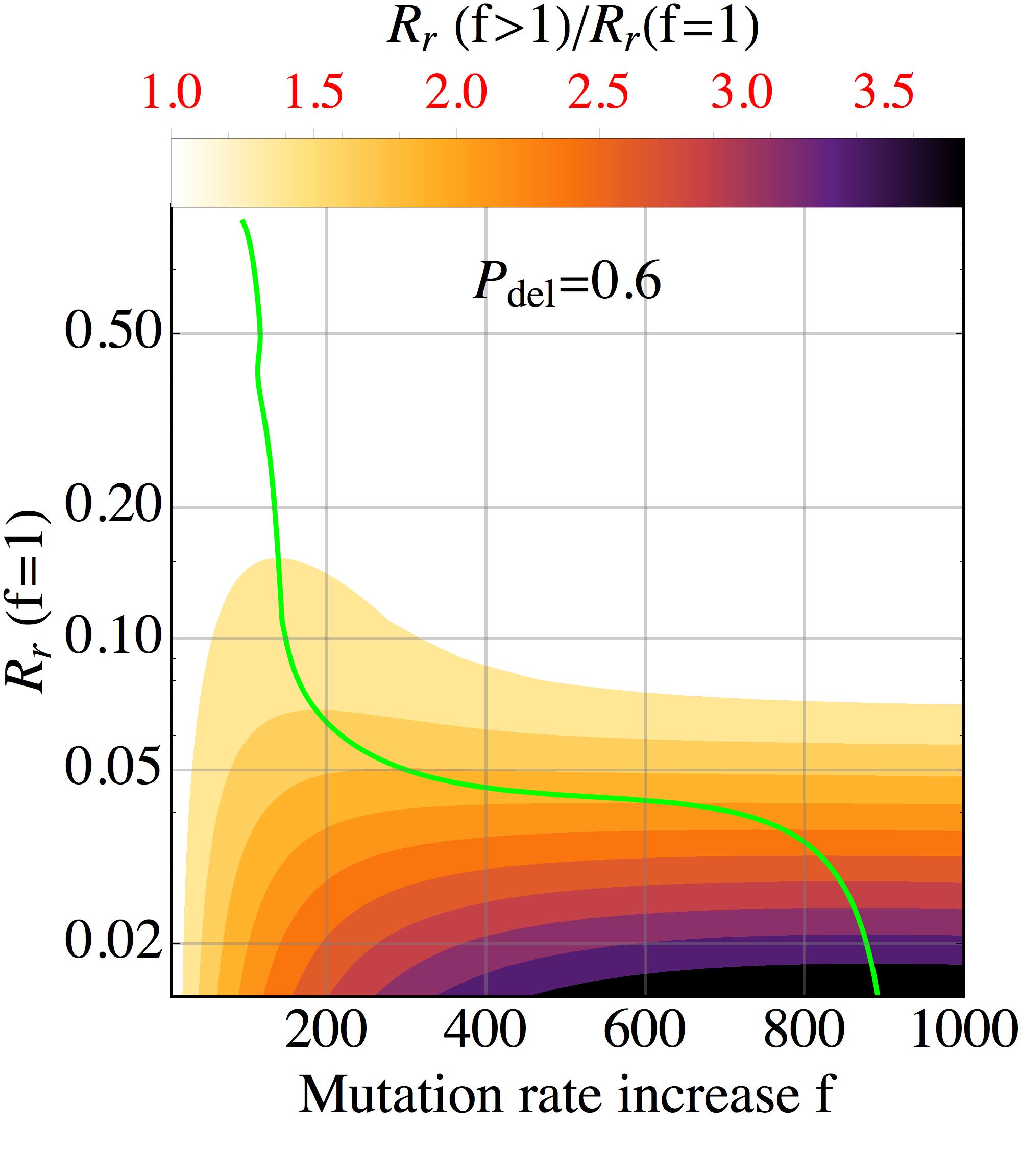} 
			\caption{}
			\label{subfig:pdel06}
		\end{subfigure}
	\end{center}
	
	\caption{\footnotesize{The sensitivity of evolution success to hypermutation ($R_r(f>1)/R_r(f=1)$) is negatively correlated with the baseline-mutation rate evolutionary advantage $R_r(f=1)$ of the resistant mutant and is optimized at a mutation rate increase ($f$) that depends on the extent of advantage, diminishing upon further increases in this rate. (a) $R_r(f>1)/R_r(f=1)$ versus $R_r(f=1)$ curves for individual parameters $\vec{\rho}$ affecting the resistant mutant's evolutionary advantage at $f=150$. (b): averages of the single-$\rho_i$ curves in the appropriate ranges put through a low pass filter for smoothness shown at multiple values of $f$ (on a log-linear scale, for clarity of resolution between different mutation rates). The dashed black line (parity in $R_r$ at baseline and elevated $f$) indicates the point of no benefit from hypermutation; initial increases in $f$ yield significant benefit at low-advantage conditions, which decreases and eventually becomes negligible at high-advantage conditions ($R_r(f=1)\rightarrow 1$). At very high $f$ (here $f\gtrsim 700$) even low-advantage mutants experience diminishing benefit from hypermutation. (c) and (d): $R_r(f>1)/R_r(f=1)$ contours corresponding to plot (b) in $f-R_r(f=1)$ space. The green curve shows the optimal (i.e. yielding highest $R_r(f>1)/R_r(f=1)$) mutation rate increase factor $f$ as a function of $R_r(f=1)$. The probability of deleterious mutations was set at $P_{del}=0.9$ for plots (a)-(c) and at $P_{del}=0.6$ in plot (d). When held constant, $\vec{\rho}$ parameters were set at $g_r/g_{wt}=3$, $K/x_{tot}(t=0)=10^2$,$p_{wt,r}=0.01$, and $x_r(t=0)=0$. Wildtype E. coli growth was set at $g_{wt}=0.34\text{ h}^{-1}$ and $\mu_{bl}=2\times 10^{-10}\times N_g$, with $N_g$ the size of the E. coli genome.}}
	\label{fig:EvAdvPlots}
\end{figure*}

The ability of a population to survive evolutionary pressure depends on the extent to which resistant phenotypes come to dominate it and withstand potential subsequent applications of the stressor (e.g. in a serial dilutions experiment). To understand how mutation rate affects this we consider how the stationary-state  ratio of resistant cells in the population, $x_r/x_{tot}$, at elevated mutation rates compares with this ratio if all mutations in the system were restricted to occur at the baseline rate $\mu_{bl}$,
\begin{equation}
\frac{R_{r}(\mu>\mu_{bl},\vec{\rho})}{R_{r}(\mu_{bl},\vec{\rho})}\equiv\frac{\left(x_{r}(\vec{\rho})/x_{tot}(\vec{\rho})\right)_{\mu>\mu_{bl}}}{\left(x_{r}(\vec{\rho})/x_{tot}(\vec{\rho})\right)_{\mu=\mu_{bl}}}\label{eq:RrRatio}
\end{equation}
as a function of the main parameters that arise in our model, $\vec{\rho}=\left(g_r,K,p_{wt,r},x_{r}(t=0)/x_{wt}(t=0)\right)$. The ratio (\ref{eq:RrRatio}) represents the extent to which a particular (elevated) mutation rate is able to drive a successful evolutionary outcome: a larger proportion of resistant cells in the stationary-state distribution. We refer to this as the \textit{sensitivity of evolutionary success to the mutation rate}. In Fig.~\ref{subfig:singleRr} we show how this sensitivity correlates with our measure of baseline ``evolutionary advantage'' of the resistant mutant, $R_{r}(\mu_{bl},\vec{\rho})$, as different components of $\vec{\rho}$ are varied\footnote{Since we consider here deterministic dynamics, the ratio $R_{r}(\mu,\vec{\rho})$ directly projects $\vec{\rho}$ to the stationary state and should therefore be viewed as both a final outcome (at stationary-state) as well as an indicator of the evolutionary advantage conferred by the system's intrinsic and extrinsic conditions.}. For each of these model parameters we see that at some fixed mutation rate (show in the plot is a 150-fold increase over the baseline) sensitivity to the elevation in mutation rate is highest at low (but positive) levels of the resistant mutant's evolutionary advantage; it decreases and eventually becomes negligible ($\approx 1$) as its evolutionary advantage increases.

Before proceeding in our analysis, we consider how the different parameters in $\vec{\rho}$ are in fact indicative of evolutionary advantage: while it is clear how a higher growth rate $g_r$, a larger initial proportion $x_{r}(t=0)/x_{wt}(t=0)$, and a higher non-deleterious mutation rate lead to advantageous conditions for the resistant mutant to increase its proportions in the population, it is perhaps less obvious why a higher resource capacity produces an advantage specifically for the resistant mutant given that resource utilization is uniform in our model among the two phenotypes. The reason for this is that while the resource capacity appears \textit{a priori} to be a non-selective environmental stressor, due to the exponential growth phase involved in the evolution of the system (\ref{eq:gensystem}), higher resource capacity puts off the time of resource saturation, thus compounding the advantage enjoyed by phenotypes with higher growth rate $g_k$.


By averaging over individual-$\rho_i$ interpolations (Fig.~\ref{subfig:aveRr}) and varying $f$ (Fig.~\ref{subfig:pdel09} and~\ref{subfig:pdel06}) we observe that the largest impact of the presence of elevated mutation rates in the population is under parameter combinations that, due to any one or multiple advantage-determining parameters, result in the resistant phenotype having a weak advantage. In these circumstances the evolutionary advantage of the resistant cells may be insufficient to establish these populations in high proportions due to competition for limited resources, and certain increases in the mutation rate may thus be critical for adaptation, even at the cost of increased deleterious mutations. When initial conditions confer a high advantage on the resistant phenotype, mutation rate increases offer negligible to negative benefit. The high growth rate of these populations and hence frequent cell divisions imply that increases in their mutation rate also drive approximately-exponentially increases in deleterious mutations, and that when a strong advantage exists the baseline-mutation phenotype will thus rise to fixation faster than its hypermutant counterpart. 

As shown in the contour plots of Fig.~\ref{subfig:pdel09} ($P_{del}=0.9$) and~\ref{subfig:pdel06} ($P_{del}=0.6$), for any level of resistant mutant evolutionary advantage, there exists an optimal mutation rate (green curves) yielding the highest proportion of resistant cells. Increasing the mutation rate up to this rate provides substantial benefit for lower-advantage mutants, and further increases lead to diminishing (albeit more gradually) returns due to the tradeoff with an increased loss caused by deleterious mutations. We see (Fig.~\ref{subfig:pdel09} compared to~\ref{subfig:pdel06}) that the level of evolutionary advantage past which there is no gain from hypermutation is fairly robust to variations in the rate of deleterious mutations ($f\mu_{bl}P_{del}$), but a lower $P_{del}$ extends the range of mutation rates conferring benefit, as in that case there is little loss to deleterious mutations even at high $f$.	

\section{Effect of selective pressure on mutation rate sensitivity and on genetic hitchhiking}\label{sec:inhibitioneffects}
In this section we focus on the effect of selective pressure in the form of an antibiotic that inhibits bacterial growth and quantify how the extent of selective pressure -- different antibiotic conentrations -- affects the sensitivity of evolutionary outcome to the mutation rate. The effect of antibiotic concentration on this quantity arises from the respective dependences of the phenotypes' growth rates on this concentration. Motivated by work~\cite{rodrigues2016biophysical} on the response of E. coli to variations in the dosage of trimethoprim, a competitive inhibitor of dihydrofolate reductase, we assume a hyperbolic decay functional dependence for the growth rate $g$ on the inhibitor concentration $[I]$
\begin{equation}
g\left([I]\right)=\frac{g_{0}}{1+[I]/g_{I}},\label{eq:biophysicalfitness}
\end{equation}
where $g_{0}$ is the growth rate in the absence of an inhibitor and $g_{I}$ controls the extent to which the population may grow in the presence of the inhibitor. In~\cite{rodrigues2016biophysical} this functional dependence, with $g_0$ and $g_I$ given explicitly as functions of various protein biophysical and cellular properties, was shown to agree with experimental measurements for several mutant phenotypes over a range of $[I]$, and similar methods can in principle be used to derive $g_0$ and $g_I$ from biophysical principles for a wider range of biologically-relevant scenarios.

\begin{figure}[h!]
	
	\centering
	\begin{subfigure}[t]{0.45\linewidth}
		\includegraphics[width=1\linewidth]{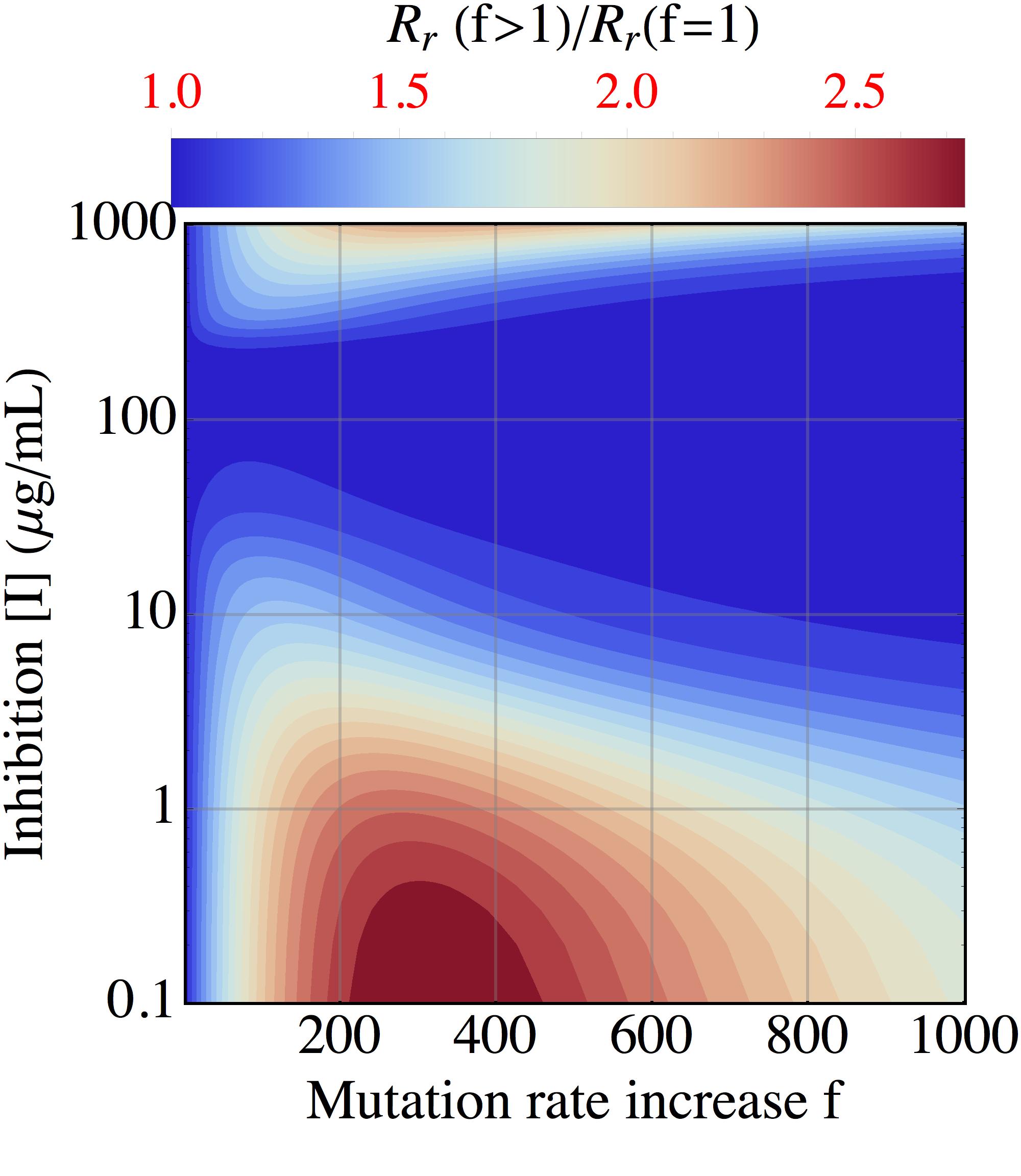} 
		\caption{}
		\label{subfig:RrInhibition}
	\end{subfigure}
	\begin{subfigure}[t]{0.45\linewidth}
		\includegraphics[width=1\linewidth]{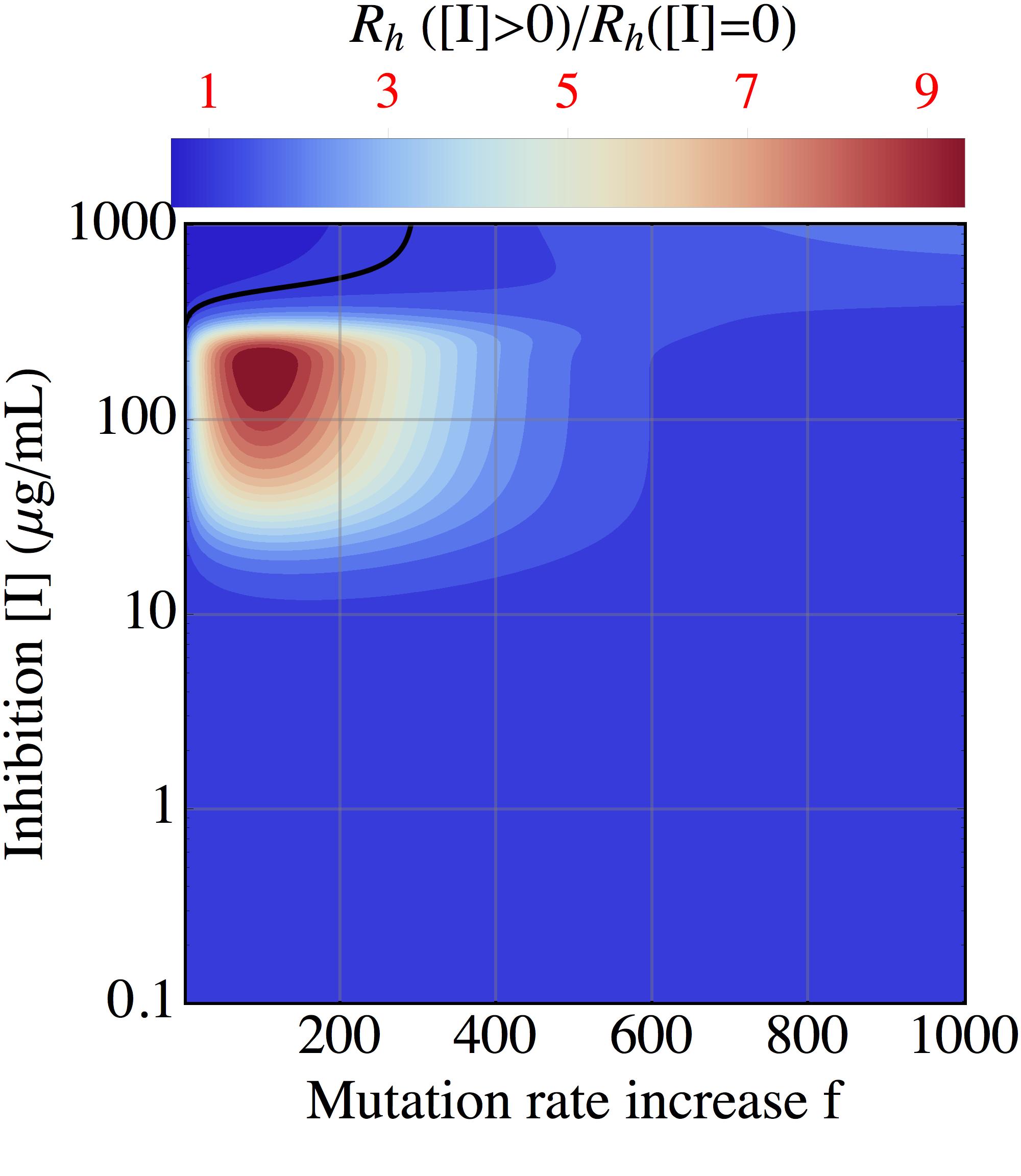} 
		\caption{}
		\label{subfig:RhyperInhibition}
	\end{subfigure}
	\caption{\footnotesize{(a): Hypermutation has strong impact on resistance fixation at low and at very high levels of inhibition that is optimized at a mutation rate that depends on the inhibition level. (b): genetic hitchhiking on resistant mutations is most pronounced in intermediate levels of inhibition. The black contour ($=1$) indicates no hitchhiking on resistant mutations. Parameters were set at $g_{0,r}=g_{0,wt}=0.34 \text{ h}^{-1}$, $g_{r,I}=5g_{wt,I}$ where $g_{wt,I}=3.6 \text{ }\mu \text{g}/\text{mL}$, $P_{del}=0.9$; and $K$, $p_{wt,r}$, and $x_r(t=0)$ as in Fig.~\ref{fig:EvAdvPlots}.}}
	\label{fig:InhibitionPlots}
\end{figure}

\begin{figure}[h!]
	
	\centering
	
		\includegraphics[width=1\linewidth]{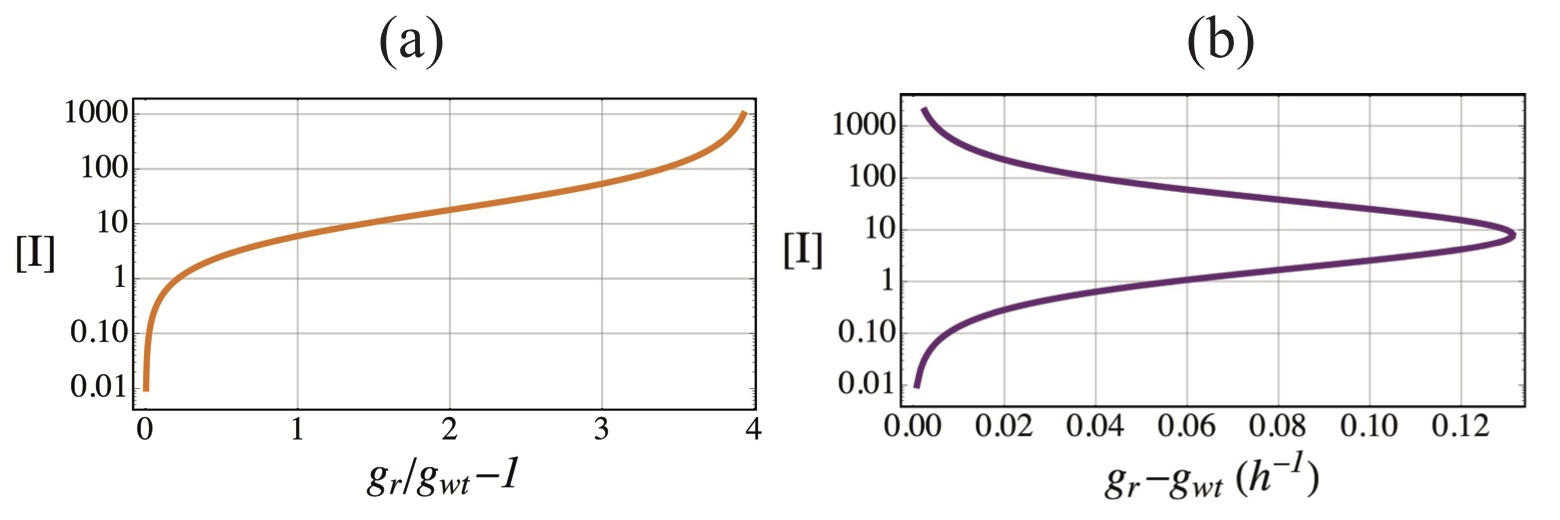} 
	\caption{\footnotesize{While the instrinsic selection coefficient $g_r/g_{wt}-1$ monotonically increases as a function of inhibition (a), the difference in growth rates $g_r-g_{wt}$ is maximized at intermediate levels of inhibition (b).}}
	\label{fig:GRatioDiffPlots}
\end{figure}

By computing the sensitivity to mutation rate, Eqn. (\ref{eq:RrRatio}), as a function of only the inhibitor with other parameters held fixed, we show (Fig.~\ref{subfig:RrInhibition}) that at low levels of inhibition, where $g_I$ carries only a small fitness advantage and mutant and wildtype growth rates $g([I])$ are similar, there is substantial benefit to be gained from hypermutation. As inhibition is increased the difference between mutant and wildtype growth increases, resulting in the resistant mutant easily increasing in proportions without much benefit from hypermutation; but at yet higher levels of inhibition the role of elevated mutation rates in determining adaptation once again becomes significant. The behavior of the (intrinsic) selection coefficient $g_r/g_{wt}-1$ is not revealing in this respect: it monotonically approaches a constant value at high $[I]$. However, the difference between $g_r$ and $g_{wt}$ peaks at an intermediate value of $[I]$ and decreases at lower and higher values of $[I]$ (Fig.~\ref{fig:GRatioDiffPlots}). While the peak does not numerically coincide with the $[I]$ concentrations yielding the lowest $R_r\left(f>1\right)/R_r\left(f=1\right)$, we note that additional parameters in $\vec{\rho}$ also affect this ratio. 

We next consider the extent to which selective pressure, the antibiotic, affects the extent of hypermutation in the population by computing the stationary-state proportion of hypermutants in the population when an inhibitor is applied (resistant cells have a positive evolutionary advantage) relative to when no inhibition is present (neutral selection),
\begin{equation}
\frac{R_{h}([I]>0,\vec{\rho})}{R_{h}([I]=0,\vec{\rho})}\equiv\frac{\left(x_{h}(\vec{\rho})/x_{tot}(\vec{\rho})\right)_{[I]>0}}{\left(x_{h}(\vec{\rho})/x_{tot}(\vec{\rho})\right)_{[I]=0}}.
\end{equation}
as a function of the inhibition (fig.~\ref{subfig:RrInhibition}). 
Fig.~\ref{subfig:RhyperInhibition} shows contours of $R_h\left([I]>0\right)/R_h\left([I]=0\right)$ in a two-dimensional space of $f$ and $[I]$.  We find (Fig.~\ref{subfig:RhyperInhibition}) that genetic hitchhiking on resistant mutations as measured by $R_{h}$ is most pronounced in a $f-[I]$ phase space that up to intermediate mutation rate increases is approximately complementary to that in which hypermutation has the most pronounced beneficial effects. This effect can be explained by noting that at low inhibition, where the resistant mutant does not have significant advantage over wildtype, the acquisition of such mutations does not drastically increase the growth rate of hypermutant cells; on the other hand, when resistant mutations are highly advantageous (high inhibition), the baseline-mutation resistant mutant rises to fixation largely unaided by hypermutation, which under finite resources limits the growth potential of other subpopulations (resistant hypermutants). We note that the range of mutation rates at which we observe hitchhiking to be strongest is in keeping with experimental observations (see~\cite{hall2006hypermutable} for a review and ~\cite{robert2018mutation} for additional recent data) of a $\mathcal{O}\left(10^1-10^2\right)$ increase over baseline in E. coli clinical isolates (with some data pointing to a nearly $\mathcal{O}\left(10^3\right)$ in certain cases~\cite{pang1985identification}).

\section{Discussion}\label{sec:discussion}

In obtaining the results presented here we assumed deterministic dynamics. While mutations typically arise randomly and can introduce a large degree of stochasticity into the dynamics, deterministic evolution can provide important insights into processes with varying degrees of stochasticity: large populations are expected to sample a large extent of the available mutational phase space (with infinite populations sampling every possible configuration, or genotype), and experimental work~\cite{toprak2012evolutionary} on evolutionary pathways in E. coli to drug resistance found similar mutational trajectories across populations evolved in parallel. Our deterministic results, moreover, suggest that stochastic fluctuations in the mutation rate can have an outsized effect on the stationary state of the system under a broad range of conditions that suppress the evolutionary advantage of emergent resistant populations.  Knowledge of the effects of these conditions in conjunction with a quantitative understanding of how changes in a controllable selective pressure, such as we modeled here in the case of a growth inhibitor, are crucial for forming informed predictions on how variations in this main driving force of adaptation affect the dynamics of complex, high-dimensional systems and on how to best minimize the effects of stochastic fluctuations to establish a desired evolutionary outcome, such as a clinical antibiotic protocol minimizing the risk of resistance evolution.

\vspace{0.4in}

\subsubsection*{Acknowledgments}
We are grateful to Jo\~ao Rodrigues for providing data on E. coli growth curves and to Michael Manhart for helpful discussions. We acknowledge support from NIGMS of the National Institutes of Health under award numbers 1R01GM124044-01 and 5R01GM068670-14.

\bibliographystyle{utphys}

\providecommand{\href}[2]{#2}\begingroup\raggedright\endgroup

\vspace{0.1in}

\appendix

\section{Evolution equations}\label{sec:eveqns}
We consider a population subdivided into wildtype with growth rate $g_{wt}$ and baseline mutation rate $\mu_{bl}$, a resistant phenotype with this baseline mutation rate and growth rate $g_{r}$, and corresponding phenotypes of equal growth rate but increased mutation rate $f\mu_{bl}$, $f>1$. Population levels at time $t$ are given by $x_{wt,1}(t)$, $x_{r,1}(t)$, $x_{wt,f}(t)$, and $x_{r,f}(t)$, respectively. Transitions (mutations) between the subpopulations are permitted in accordance with the schematic Fig.~\ref{fig:schematic}. We assume limited resources set by an environmental carrying capacity $K$, so that the subpopulation levels thus evolve in time according to the equations

\onecolumngrid
\begin{equation}
	\begin{cases}
		\dot{x}_{wt,1}(t)=\left(1-\frac{x_{tot}(t)}{K}\right)\left[\left(1-\mu_{bl}\left(p_{r, wt}+P_{del}\right)-r_{\mu}\right)g_{wt}x_{wt,1}(t)+\mu_{bl}p_{r,wt}g_{r}x_{dr,1}(t)+r_{\mu}g_{wt}x_{wt,f}(t)\right]\\
		\dot{x}_{wt,f}(t)=\left(1-\frac{x_{tot}(t)}{K}\right)\left[\left(1-f\times \mu_{bl}\left(p_{r,wt}+P_{del}\right)-r_{\mu}\right)g_{wt}x_{wt,f}(t)+f\times \mu_{bl}p_{r,wt}g_{r}x_{r,f}(t)+r_{\mu}g_{wt}x_{wt,1}(t)\right]\\
		\dot{x}_{r,1}(t)=\left(1-\frac{x_{tot}(t)}{K}\right)\left[\left(1-\mu_{bl}\left(p_{r,wt}+P_{del}\right)-r_{\mu}\right)g_{r}x_{r,1}(t)+\mu_{bl}p_{r,wt}g_{wt}x_{wt,1}(t)+r_{\mu}g_{r}x_{r,f}(t)\right]\\
		\dot{x}_{r,f}(t)=\left(1-\frac{x_{tot}(t)}{K}\right)\left[\left(1-f\times \mu_{bl}\left(p_{r,wt}+P_{del}\right)-r_{\mu}\right)g_{r}x_{r,f}(t)+f\times \mu_{bl}p_{r,wt}g_{wt}x_{wt,f}(t)+r_{\mu}g_{r}x_{r,1}(t)\right]
	\end{cases}\label{eq:fourtypedynamics}
\end{equation}
\twocolumngrid
\noindent where $r_{\mu}$ is the rate of mutation from a baseline-mutation rate ($f=1$) phenotype to a $f>1$ phenotype (rate at which mutations leading to elevated mutation rates $f\mu_{bl}$ occur) and its reverse (assumed to be equal),  $p_{r,wt}\equiv p_{wt\rightarrow r}=p_{r\rightarrow wt}$ is the probability of mutation from wildtype to the resistant phenotype and backward, $P_{del}$ is the probability of mutation to deleterious phenotypes, $x_{tot}=x_{wt,1}+x_{wt,1f}+x_{r,1}+x_{r,f}$.

In order to compute the relative advantage or disadvantage conferred by hypermutation on the fixation of drug resistant subpopulations (Eq.~\ref{eq:RrRatio}) we numerically compute the ratio of resistant mutants (combined non-hypermutant and hypermutant types) in the total population at $2\leq f\leq 1000$ to the ratio that would result if no hypermutations were allowed in the system, i.e. if we set $r_{\mu}=0$ and consider only phenotypes $x_{wt,1}$ and $x_{r,1}$. When computing these quantities for a system with an initial distribution of hypermutants of either phenotype, we assume that the $r_{\mu}=0$ system has a corresponding distribution in which $$x_{i,r_{\mu}=0}\left(t=0\right)=x_{i,1,r_{\mu}\neq 0}\left(t=0\right)+x_{i,f,r_{\mu}\neq0}\left(t=0\right),$$ with $i$ representing either wildtype or the resistant phenotype.

In the figures shown in the main text and in the Supplemental Material $\mu_{bl}$ was set at $2\times 10^{-10}\times N_{\text{genome}}$ per generation per cell~\cite{lee2012rate} where $N_{\text{genome}}=4.64\times 10^6$ is the number of basepairs in the E. coli genome. In the results shown in the main text, when the population is not purely wildtype at $t=0$, the proportion of hypermutants chosen is assumed to be distributed proportionally amongst the wildtype and resistant populations.

\section{Rate of acquisition $r_{\mu}$ of increased mutation rate and initial proportion of hypermutants}\label{sec:rmut}

To estimate a biologically reasonable $r_{\mu}$ we consider a simple
system consisting of a wildtype $f=1$ phenotype and a wildtype $f>1$,
both with fitness $g_{wt}$, which can mutate into each other with
rate $r_{\mu}$:
\onecolumngrid
\begin{equation}
	\begin{cases}
		\dot{x}_{wt,1}(t)=\left(1-\frac{x_{wt,1}(t)+x_{wt,f}(t)}{K}\right)\left[\left(1-r_{\mu}\right)g_{wt}x_{wt,1}(t)+r_{\mu}g_{wt}x_{wt,f}(t)\right]\\
		\dot{x}_{wt,f}(t)=\left(1-\frac{x_{wt,1}(t)+x_{wt,f}(t)}{K}\right)\left[\left(1-r_{\mu}\right)g_{wt}x_{wt,f}(t)+r_{\mu}g_{wt}x_{wt,1}(t)\right]
	\end{cases}
	\label{eq:2typesys}
\end{equation}
\twocolumngrid
The steady-state (stationary distribution) proportion of hypermutants
in the total population will be given by 
\begin{equation}
	R=\frac{x_{wt,f}(\tau)}{x_{wt,1}(\tau)+x_{wt,f}(\tau)}\label{eq:hyperratio}
\end{equation}
at time $\tau$ after resources have been saturated\footnote{Since our goal in the analysis of this section is to obtain a first-order estimate for $r_{\mu}$ that is independent of the precise mutation rate we omit the potential effect of deleterious mutations here via $P_{del}$, as the magnitude of this effect will depend on the actual mutation frequency $f\mu_{bl}$ of the hypermutators.}. Hypermutation can be caused by various mechanisms; studies focused on pathogenic E. coli have found comparatively high ($>1\%$) proportions of mutators in bacterial isolates ($3.6\%$ in~\cite{jyssum1960observations} and $1.9\%$ in~\cite{leclerc1996high}); a separate study that looked specifically for MMR deactivation in E. coli found a much lower proportion ($0.24\%$) when both commensal and pathogenic E. coli were included~\cite{gross1981incidence}. A later study~\cite{matic1997highly} found, however, that when other sources of hypermutation were included besides MMR, E. coli cells exhibiting increased mutation rates -- of up to two orders of magnitude from the baseline mutation rate -- constituted as much as $14\%$ of the total population, most being mild mutators, with both commensal and pathogenic strains included in the study. The highest mutation rates were found to correspond to MMR deficiencies, with lower increases due to other mechanisms. Note that since $r_{\mu}$ is a neutral-selection rate, studies of mutator proportions that were conducted under conditions of adaptive evolution will likely overestimate this parameter and we therefore restrict our data to studies of natural isolates, noting that even in those cases adaptive evolution in the recent past may have taken place.

\begin{figure}[h]
	\centering
	\includegraphics[width=0.45\textwidth]{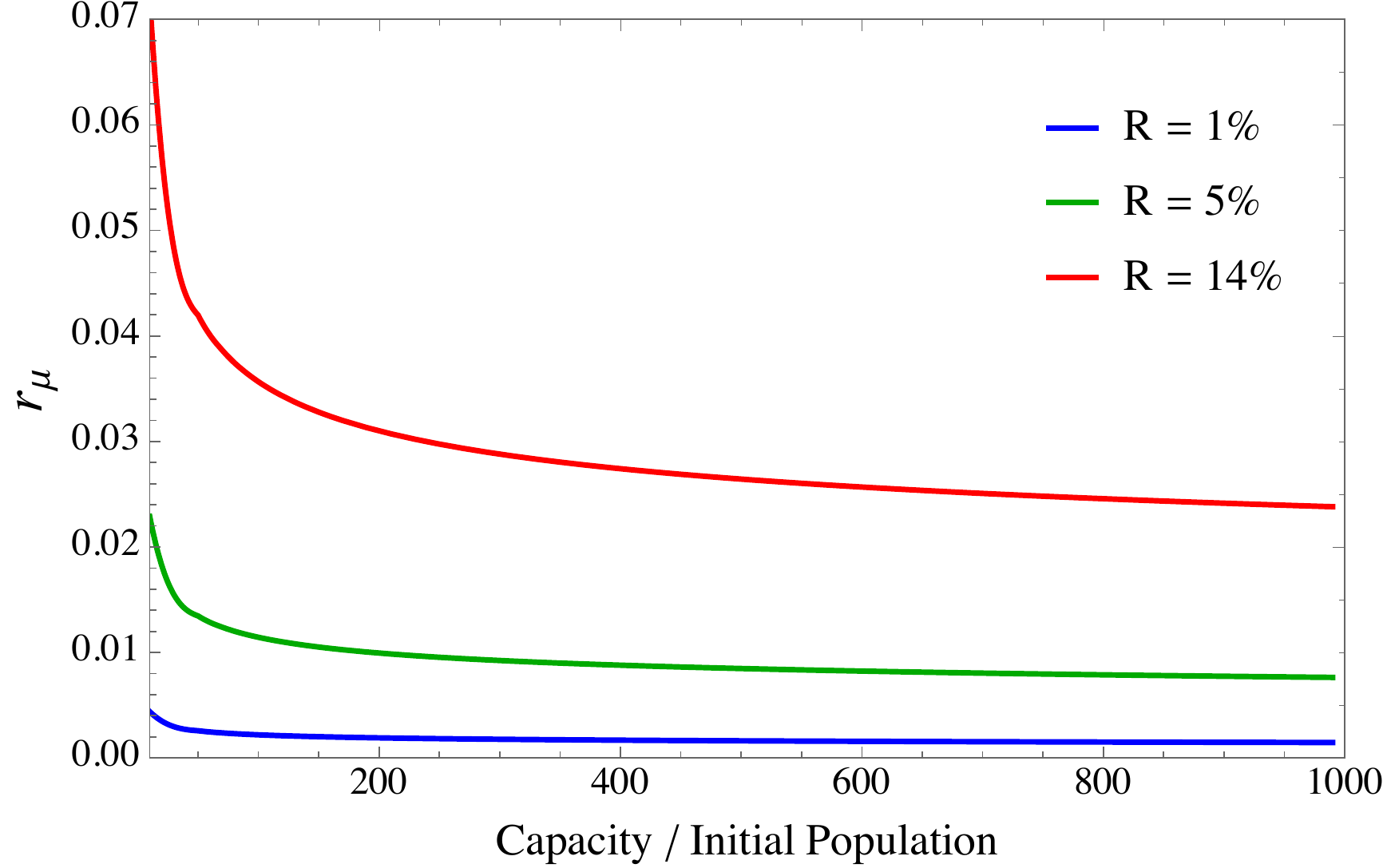}
	\caption{\small{Computed values of $r_{\mu}$ from the system (\ref{eq:2typesys}) for different expected steady-state proportions of resistant cells as the availability of resources is varied. Across a wide range of carrying capacities $r_{\mu}$ only varies from $\sim 0.5\%$ to $\sim 1.5\%$. Plots in the main text employ $r_{\mu}=0.25\%$, corresponding to an initial hypermutant population of $1\%$ of the total population.}}
	\label{fig:rmutvscap}
\end{figure}

We compute which $r_{\mu}$ values yield the stationary distribution
ratio (\ref{eq:hyperratio}) for different carrying capacities  by taking $g_{wt}$ as in the main text to be 0.34 $\text{h}^{-1}$ under no inhibition. The results for different values of $R$ are shown in Fig.~\ref{fig:rmutvscap}. Since we consider a uniform distribution\footnote{A non-uniform distribution can be incorporated by multiplying $r_{\mu}$ by a probability distribution that depends on $f$ - as this adds additional degrees of freedom to the model we avoid doing so here.} of mutation rate increase factors $f$ and the $14\%$ figure is heavily tipped towards mild mutators, using this figure will likely overestimate the mutation rate $r_{\mu}$ in our model for higher values of $f$. For the purpose of the plots in the main text we set on the lower end, at $r_{\mu}=0.25\%$ and an initial proportion of $1\%$ hypermutating cells in the population.

\clearpage



\section*{Supplementary material}

# Supplemental Material for "Mutation rate variability as a driving force in adaptive evolution"
Dalit Engelhardt and Eugene I. Shakhnovich

Department of Chemistry and Chemical Biology, Harvard University, Cambridge, MA 02138

## S1 Results for a spectrum of $r_\mu$ and $P_{del}$

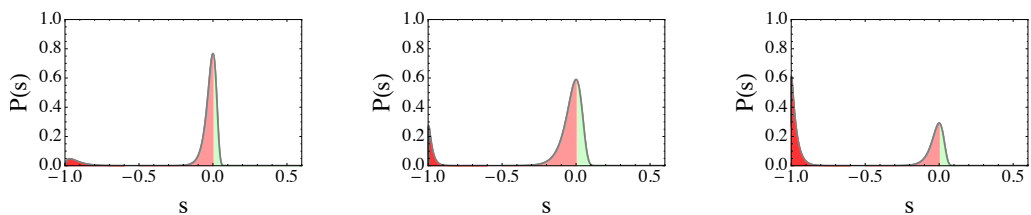

Figure S1: Sample bimodal distributions of fitness effects shown in terms of the probability of mutation as a function of the selection coefficient, $s \equiv g_{\text{mut}}/g_{wt} - 1$, with the green and light red regions indicating mutations with beneficial and mildly deleterious effects, respectively. The darker red region indicates strongly deleterious and lethal mutations.

Experimental findings on distributions of fitness effects (DFEs) across a variety of organisms and conditions point to a probability peak around neutral (wildtype) fitness that is skewed toward mildly to somewhat deleterious mutations (Fig S1). The proportion of lethal or near-lethal mutations varies greatly in the literature and exhibits a dependence on both the organism studied and the experimental conditions: earlier studies (e.g. [1]) found a bimodal distribution to be typical, whereas a recent study has observed far fewer lethal mutations in E. coli, consituting only about 1% of all mutations recorded [2]. In view of these differing data and the possibility that ambient conditions may affect the DFE, it is important to establish the extent to which our conclusions in this study are robust against variations in the balance between the rate of deleterious, neutral, and beneficial mutations. The probability of beneficial mutations was one of the main indicators of selective advantage that we explored in the main text as a component of advantage-conferring conditions $\vec{\rho}$; here we consider changes in $P_{del}$ – the probability assigned to strongly deleterious or lethal mutations in our model.

Figs. S2 and S3 show plots corresponding to Fig. 2(c)-(d) and Fig. 3 in the main text for different combinations of $P_{del}$ and $r_\mu$ with corresponding choices of initial hypermutant proportion, $x_{h,0}/x_{tot,0}$ based on the analysis in the previous section. We see from Fig. S2 that the effects of increasing $P_{del}$, i.e. altering the balance of beneficial to deleterious mutations to result in a higher rate of deleterious mutations, are (1) to shift the optimal mutation rate (i.e. to more strongly constrain the range of mutation rate increases $f$ that benefit resistance fixation) to lower values for a given level of baseline-rate advantage ($R_r(f = 1)$); and (2) to increase the overall impact of hypermutation ($R_r(f > 1)/R_r(f = 1)$) for a given level of baseline-rate advantage. Importantly, however, the level of advantage at which hypermutation no longer plays a role does not significantly change over a large range of variations in $P_{del}$, indicating that our main results about the role of hypermutation as a function of evolutionary advantage are qualitatively robust against changes in the distribution of fitness effects. We see that changes in $r_\mu$ and $x_{h,0}/x_{tot,0}$ carry a significant quantitative impact on $R_r(f > 1)/R_r(f = 1)$ and that higher values of $r_\mu$ and $x_{h,0}/x_{tot,0}$ result in little to no loss in $R_r(f > 1)/R_r(f = 1)$ even at very high mutation rates. The effects due to changes in $P_{del}$ and $(r_\mu, x_{h,0}/x_{tot,0})$ carry over to the behavior of $R_r(f > 1)/R_r(f = 1)$ and $R_h([I] > 0)/R_h([I] = 0)$ as shown in Fig. S3.

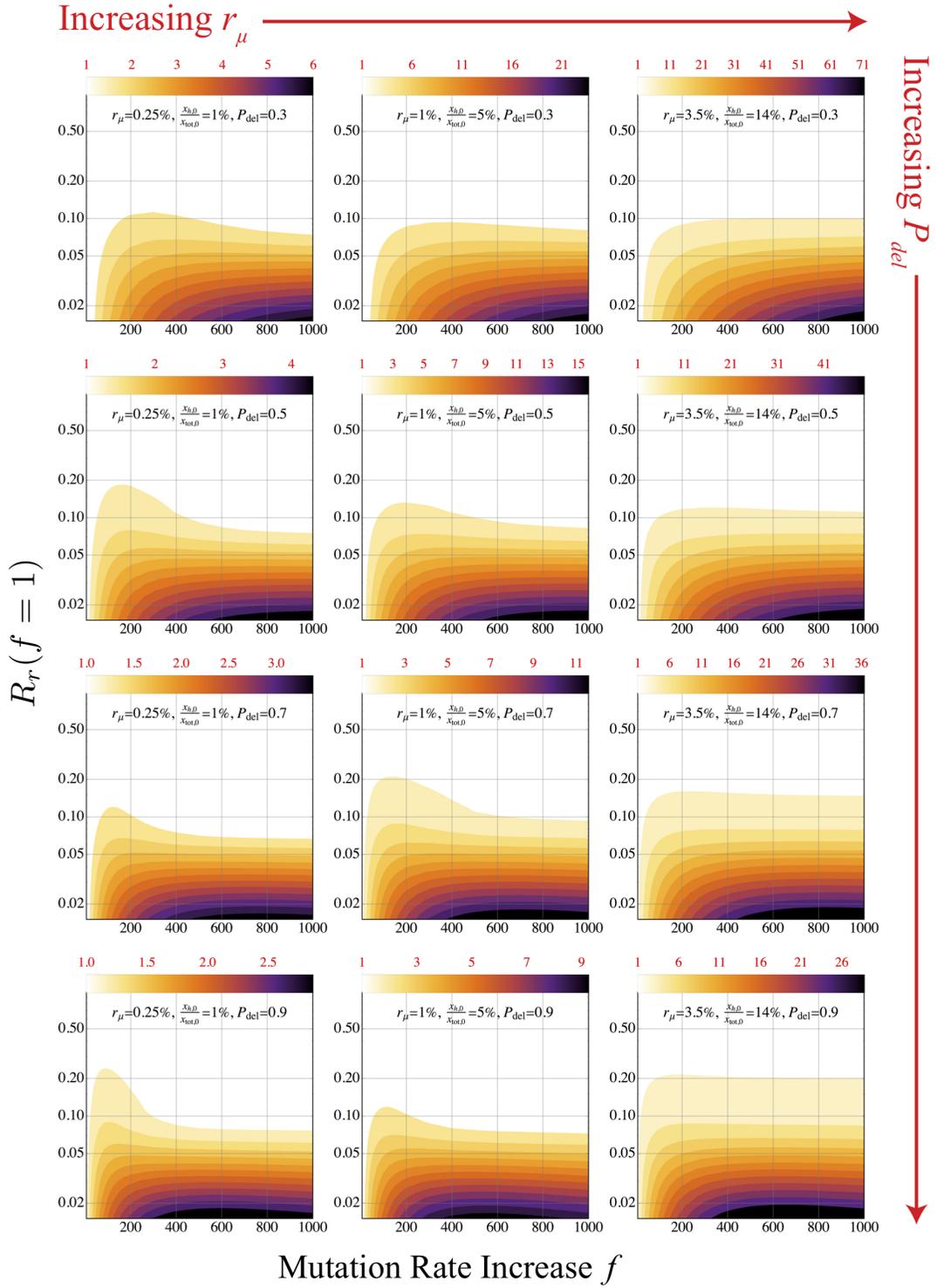

Figure S2: Contour plots of $R_r(f > 1)/R_r(f = 1)$ (Eqn. (3.1) in the main text) in $f$-$R_r(f = 1)$ space corresponding to those shown in Fig. 2(c)-(d) in the main text (with parameters noted there) at different combinations of $(r_\mu, P_{del})$ and the approximate[2] proportion of initial-state hypermutants $(x_{h,0}/x_{tot,0})$ corresponding to this choice of $r_\mu$ based on the analysis plotted in Fig. 5 in the Appendix.

[2] Since we consider a range of $K$ values in calculating $R_r(f > 1)/R_r(f = 1)$, the choice of $x_{h,0}/x_{tot,0}$ corresponding to $r_\mu$ is approximate.

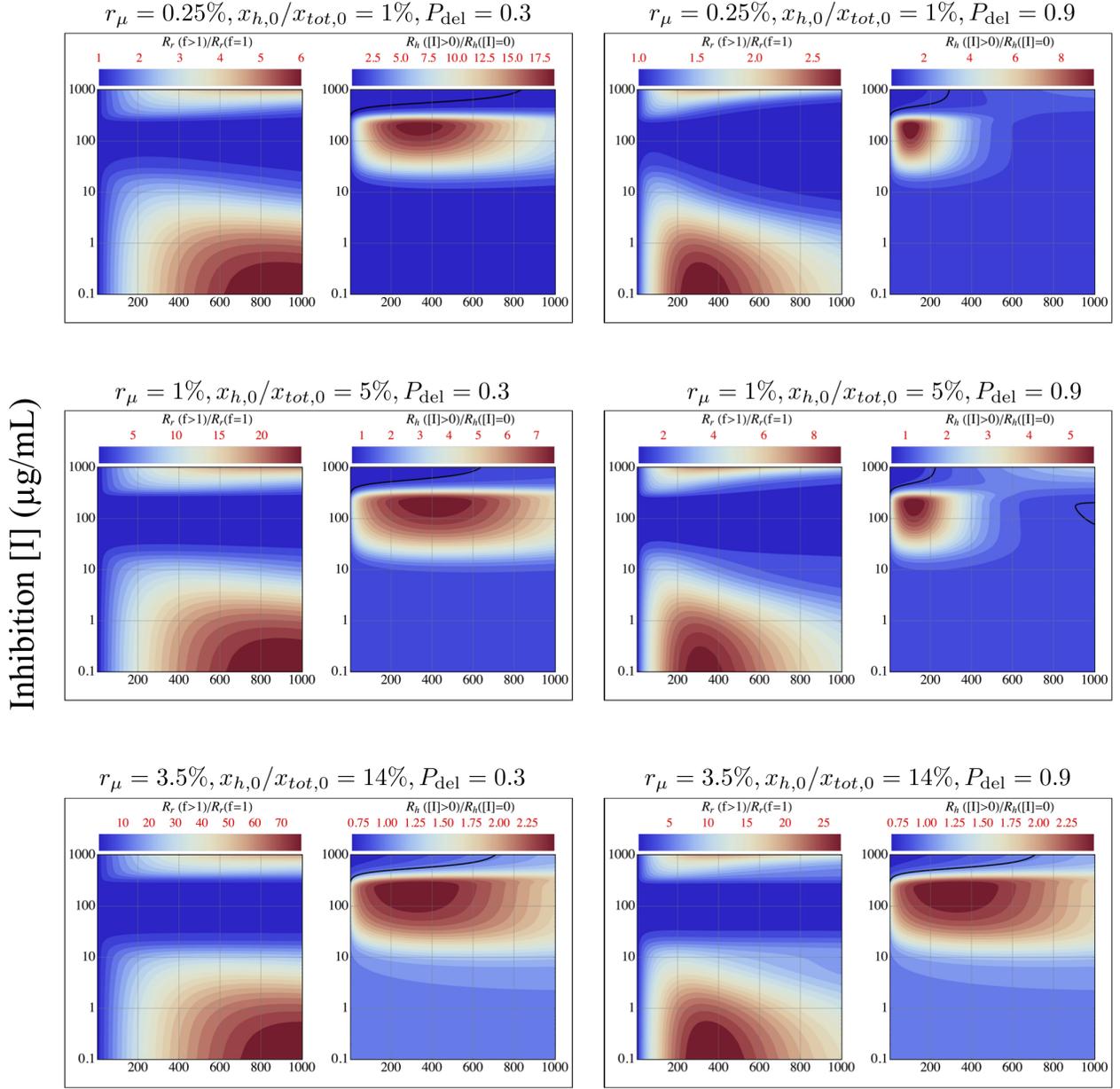

Figure S3: Side-by-side contour plots for the effects of inhibition on resistance fixation ($R_r(f>1)/R_r(f=1)$) and hypermutation fixation ($R_h([I]>0)/R_h([I]=0)$) in $f$-$[I]$ space corresponding to those shown in Fig. 3(a)-(b) in the main text (with parameters noted there) at different combinations of ($r_\mu, P_{del}$) and the approximate proportion of initial-state hypermutants ($x_{h,0}/x_{tot,0}$) corresponding to this choice of $r_\mu$ based on the analysis plotted in Fig. 5 in the Appendix. The location of $R_h([I]>0) = R_h([I]=0)$ is shown in the black curve, indicates the point of no hitchhking of hypermutants on resistant mutations.

## S2 Subpopulation fixation dynamics

Shown below are the frequencies of each of the four subpopulations in the model as a function of time at different mutation rates. In the main text, proportions were calculated from the stationary population distribution.

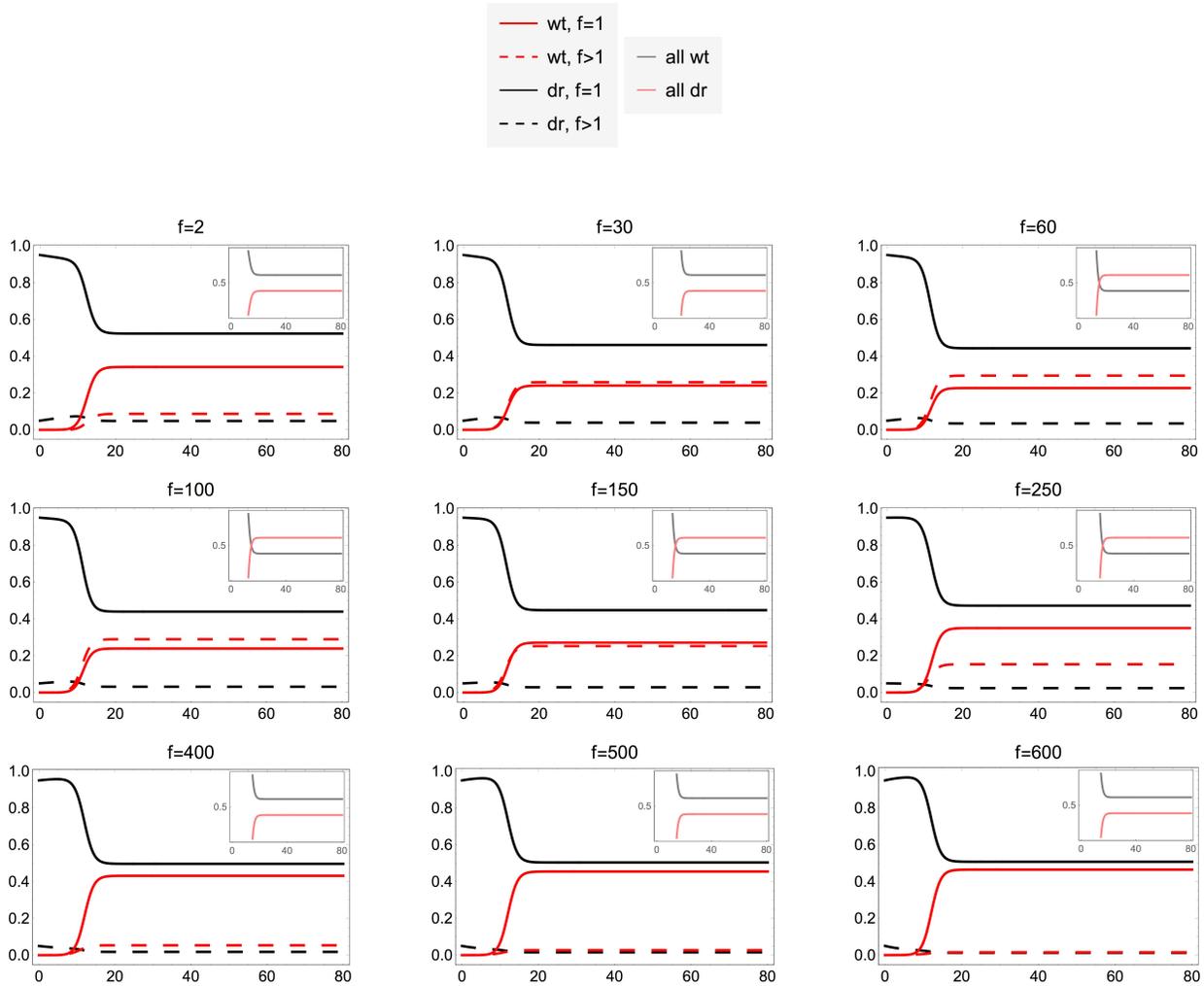

Figure S4: Frequencies (proportions in total population) over time of each of the four subpopulations in the model: baseline-mutation rate wildtype (solid red), baseline-mutation rate resistant (solid black), hypermutant wildtype (dashed red), and hypermutant resistant (dashed black). The inset plots show the frequencies of all wildtype (hypermutant and baseline combined) cells (gray) and all resistant cells (light red). Intermediate mutation rate increases ($f = 60, 100, 150, 250$) in this case result in the resistant mutant becoming dominant by the time that resource saturation and hence a stationary-state distribution are reached, whereas lower and higher mutation rate increases have a wildtype-dominated stationary-state distribution. Model parameters were set at $g_r = 3.5 \times g_{wt}$ with $g_{wt} = 0.34$ h$^{-1}$, $x_{tot}(t=0) = 10^6$, $K = 10^8$, $P_{wt \to r} = 0.01$, $x_r(t=0) = 0$, $x_h(t=0) = 5\%$ (all wildtype hypermutants), $r_\mu = 1\%$, and $P_{del} = 0.9$.



\end{document}